%% file: driverI2.tex
\documentclass{tcibook}
\usepackage{fancyhea}
\usepackage{work}
\usepackage{bm}       
\usepackage{graphicx}
\usepackage{hyperref}      
\usepackage{color}

\input workshopsymbols.tex      

\begin{document}

\def\bibname{References}
\bibliographystyle{plain}

\raggedbottom

\pagenumbering{roman}

\parindent=0pt
\parskip=8pt
\setlength{\evensidemargin}{0pt}
\setlength{\oddsidemargin}{0pt}
\setlength{\marginparsep}{0.0in}
\setlength{\marginparwidth}{0.0in}
\marginparpush=0pt


\pagenumbering{arabic}

\renewcommand{\chapname}{chap:intro_}
\renewcommand{\chapterdir}{.}
\renewcommand{\arraystretch}{1.25}
\addtolength{\arraycolsep}{-3pt}


\input wgreport.tex



\end{document}

%% file: workshopsymbols.tex
\newcommand{\nc}{\newcommand}  

\def\beq{\begin{equation}}
\def\eeq#1{\label{#1}\end{equation}}
\def\eeqn{\end{equation}}

\newenvironment{Eqnarray}%
   {\arraycolsep 0.14em\begin{eqnarray}}{\end{eqnarray}}
\def\beqa{\begin{Eqnarray}}
\def\eeqa#1{\label{#1}\end{Eqnarray}}
\def\eeqan{\end{Eqnarray}}

\nc{\ra}{\rightarrow}  
\nc{\slsh}{\slash\hspace*{-0.22cm}}
\def\Re{{\cal R \mskip-4mu \lower.1ex \hbox{\it e}\,}}
\def\Im{{\cal I \mskip-5mu \lower.1ex \hbox{\it m}\,}}

\nc{\vev}[1]{ \left\langle {#1} \right\rangle }
\nc{\bra}[1]{ \langle {#1} | }
\nc{\ket}[1]{ | {#1} \rangle }
\nc{\fb}{\,{\rm fb}^{-1}}
\nc{\ev}{{\rm eV}}
\nc{\kev}{{\rm keV}}
\nc{\Mev}{{\rm MeV}}
\nc{\gev}{{\rm GeV}}
\nc{\tev}{{\rm TeV}}
\nc{\mev}{{\rm MeV}}

\def\del{\partial}
\def\Dslash{\not{\hbox{\kern-4pt $D$}}}
\def\dslash{\not{\hbox{\kern-2pt $\del$}}}
\def\pslash{\not{\hbox{\kern-2pt $p$}}}
\def\ETmiss{ \not{\hbox{\kern-4pt $E$}}_T }

\def\msb{{\bar{\ssstyle M \kern -1pt S}}}



%% file: wgreport.tex
 
\chapter{Computing Frontier: Distributed Computing and Facility Infrastructures}
\label{chap:mag}

\begin{center}\begin{boldmath}

\input authorlist.tex


\end{boldmath}\end{center}


\section{Introduction}
\label{sec:comp-intro}

The field of particle physics has become increasingly reliant on large-scale computing resources to address the challenges of analyzing large datasets, completing specialized computations and simulations, and allowing for wide-spread participation of large groups of researchers.  For a variety of reasons, these resources have become more distributed over a large geographic area, and some resources are highly specialized computing machines.

In this report for the Snowmass Computing Frontier Study, we consider several questions about distributed computing and facility infrastructures that could impact future resource requirements and research directions.  Two other efforts to understand these issues during the past year have been major resources for this report.  One was a review conducted by NERSC in November 2012 to determine HEP community computing and storage needs through 2017~\cite{bib:NERSCreport}.  Another was a panel discussion on the future of grid computing held at the Open Science Grid All-Hands Meeting in March 2013~\cite{bib:OSGpanel}.  We thank all of the participants in these discussions for their contributions.

\section{Current HEP Use of the U.S. National Computing Infrastructure}

Different computational problems in particle physics are naturally suited for different kinds of computing facilities.  In general, there are two paradigms.  One is HTC, which is implemented in standard commodity computers and can address problems that are ``embarrassingly parallel,'' i.e., those that can be computed independently with the results combined afterward.  The other is HPC, which uses supercomputers to solve large problems by distributing computational work among many processors and using specialized high-speed, low-latency networks to communicate partial results among processors during execution of the job.

Historically, HPC machines for open science have been located at specialized national centers funded by the 
Department of Energy and the National 
Science Foundation. These supercomputers were designed and built to solve large-scale computational problems, 
typically tightly coupled simulations requiring fast processors, a high-speed internal network, and sometimes a 
fast I/O subsystem. In recent years a need for HTC has quickly grown in the science community. Part of this demand has arisen from
the data-driven science (e.g. LHC data analysis) and part has from the simulation community's need to perfom 
massive numbers of ensemble runs to explore paramater space and test and validate codes. 
In response to this demand, national centers have been developing capabilities geared towards supporting the HTC paradigm.  
For example, NERSC hosts and operates systems that run HTC workflows for genomics, high energy and nuclear physics,
astronomy, and materials science. Software has also been developed to support these workflows on the largest HPC systems.
Thus, the centers are in a stronger position to support a wider variety of computing tasks, and it is possible that both HPC- and HTC- driven science can find a home at these national centers.  Some successful examples are mentioned later in this report.

The Worldwide LHC Computing Grid (WLCG) is an HTC resource that is the main computational resource used by the LHC experiments, of which ATLAS and CMS have the largest computing needs.  As the name implies, the WLCG is an example of a grid infrastructure, which is described in much more detail in Section~\ref{sec:grid}.  There are over 170 facilities connected to the WLCG, distributed over 36 countries.  Fifteen of those sites are located in the United States, and they tend to have more resources than the average WLCG site.  The WLCG is organized into a tiered hierarchy of sites, in which sites at each tier have different computational responsibilities and service levels, and thus different hardware configurations.  The Tier-0 center is at CERN; it is responsible for prompt reconstruction of detector data, some calibration and alignment tasks and keeping a custodial copy of the raw data.  There are currently twelve Tier-1 sites, which keep a second custodial copy of the raw data, reprocess older data with improved calibration and alignment constants, perform skims of large data samples, and archive simulated datasets.  Both Tier-0 and Tier-1 centers operate tape libraries and have 24/7 system support.  The remaining sites are Tier-2 sites, which host data samples for physics analysis and generate simulated datasets.  Tier-2 centers typically only have business-hours support within their time zone.  The facilities are composed of large clusters of commodity machines powered by x86-style processors, which are accessed through batch scheduling systems.

The computational problems of the LHC experiments are well matched to the structure of the WLCG.  Computations are centered around individual, statistically independent collision events, and this embarrassingly parallel regime works well for the HTC systems that the WLCG provides.  This scheme has served the LHC experiments very well.  The current resources of the WLCG are 2 MHS06 of CPU and 190 PB of disk.  CMS and ATLAS used about 300,000 cores continuously during 2012, resulting in about 2.6 billion CPU hours.  These resources, along with robust middleware and a strong effort in operations, have allowed the experiments to turn around physics results very quickly.  The workflows for the experiments, be they for data processing, calibration, simulation or user analysis, have performed as expected, and any concerns about scaling with the expected increase in resources should be able to be addressed in the course of normal operations.  There is a good window for this work during the current LHC shutdown.  It should also be noted that the specific assignments of particular workflows to particular tiers of the infrastructure is expected to evolve in the coming years, to make the most efficient use of all available resources.

As discussed below, whether the WLCG will continue to serve the needs of the LHC experiments depends very much on how WLCG capacity evolves, and how efficiently the experiments can make use of it.  This is an important question, given the anticipated growth in LHC luminosity (from $7 \times 10^{33}$ to $1 \times 10^{34}$/cm$^2$/s), event complexity due to pileup (from a typical 20 extra interactions per event in the previous LHC run to 25 in the 2015 run), and trigger rate to maintain sensitivity to the Higgs boson and new-physics signatures (from 300 Hz to perhaps 1 kHz).  However, any changes to the WLCG usage can be made in an evolutionary fashion, and the underlying paradigm of HTC should continue to work.

Because of the sheer scale of the existing WLCG resources, we anticipate that the WLCG will remain the main resource for LHC experiment computations.  However, the use of other facilities, such as those described below, should be explored to see if they can successfully perform the same computations and thus augment the LHC computing capacity.

Intensity Frontier experiments have a diverse set of needs, but in aggregate they have large data and analysis requirements. While these experiments are generally served by HTC facilities, a number of existing experiments have successfully taken advantage of HPC centers' interest in enabling data-driven scientific discovery through data-intensive HTC computing.  For example, analysis for the Daya Bay experiment was conducted at DOE's NERSC center, which also served as the Tier 1 data center.  DOE HPC centers have also supported KAMLAND, IceCube, BaBar, SNO, ALICE, ATLAS, and Palomar PTF data analysis.

National High Performance Computing centers are used and required in a number of HEP areas of research, including

\begin{itemize}
\item Lattice QCD (Energy Frontier)
\item Accelerator design and R\&D (Energy and Intensity Frontiers)
\item Data analysis and creations of synthetic maps (Cosmic Frontier)
\item N-body and hydrodynamic cosmology simulations (Cosmic Frontier)
\item Supernova modeling (Cosmic Frontier)
\end{itemize}

\begin{figure}[h]
\includegraphics[width=\textwidth]{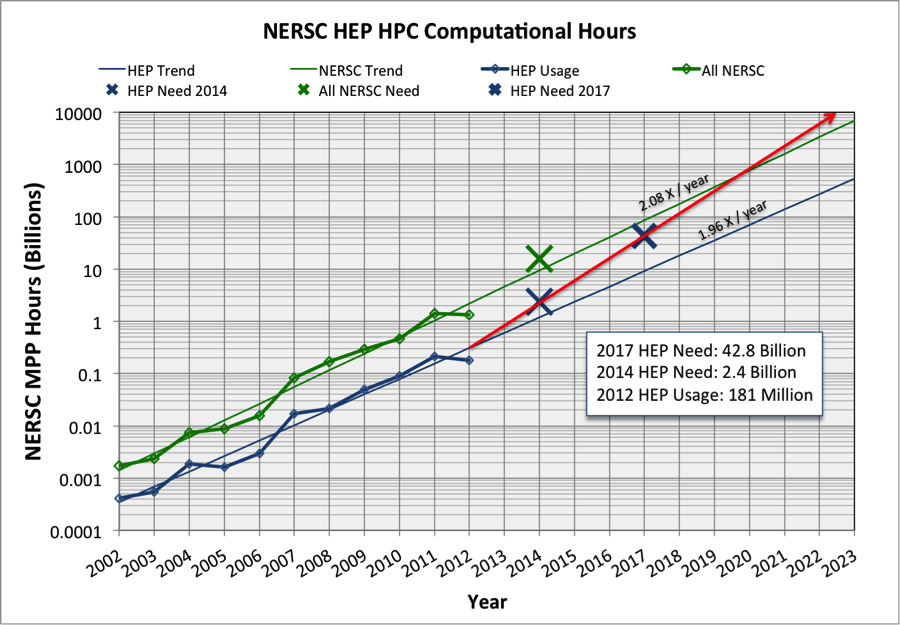}
\caption{Historical normalized computing hours used at NERSC (green line) and just for HEP projects (blue line). Results from NERSC requirements reviews with HEP scientists and DOE program managers (large blue crosses) show a need for computing greater than what will be supplied by extrapolating the trend.}
\label{fig:NERSC-Computational-Hours}
\end{figure}

A great need for HPC computing, driven by the needs of LQCD and computational cosmology but required by other fields as well, will outpace even the historical trend (see Figure \ref{fig:NERSC-Computational-Hours}), even as extrapolation of those trends becomes uncertain due to power and technology limitations.

One effort, Perturbative QCD,  that has largely relied on HTC computing to date has already started using HPC centers and expects to 
expand those efforts to be able to complete the calculation of important background and signal reactions at the
LHC.  
They have determined that it would be beneficial to make the national DOE HPC
facilities 
generally accessible to particle theorists and
experimentalists in order to enable the use of existing
calculational tools for experimental studies involving extensive
multiple runs without depending on the computer power and manpower
available to the code authors. Access to these facilities will also
allow prototyping the next generation of parallel computer programs
for QCD phenomenology and precision calculations.

\section{Future Availability of Resources}

The needed computing resources for Energy Frontier experiments are currently set by the needs of the CMS and ATLAS experiments.  Both of these experiments have had several years of running experience and have developed the tools to predict future resource needs as a function of experimental parameters such as the trigger rate, pileup distribution, event size, number of reprocessing and analysis passes per year and so forth.  These models have been shown to be reasonably predictive~\cite{bib:CHEPresources}.

It currently appears that the needs will be met for the foreseeable ($\sim$10 year) future as long as several conditions are satisfied.  So far, the funding for the WLCG has been roughly constant, allowing resource growth to continue with Moore's Law.  Experiments have been able to adjust their computing models to adapt to the available growth in resources along with the growth in data sets.  While Moore's Law does not seem to hold as well as it used to, resources should still grow over time, although not as quickly as before.  Near-constant WLCG funding will be necessary for the experiments to keep up with growing datasets and event complexity.  

Meanwhile, computing architectures are changing, and the experiments' software bases must evolve to keep up with them.  Adapting the software to take advantage of many-core and multicore processor architectures is also critical for LHC experiments to meet their computing needs.  The experiments will also need to find greater efficiencies in resource usage, for both processing and disk resources.  Currently both ATLAS and CMS distribute many datasets to their computing sites that are subsequently rarely or never used; this is then a waste of storage.  The experiments will also need to proactively pursue and take advantage of a variety of resources beyond the WLCG.  These include opportunistic resources that might be available at universities, laboratories and NSF and DOE computing centers, and paid resources that might be available through commercial clouds.  Fortunately, both CMS and ATLAS are actively pursuing many of these measures, which are an important part of the development plans underway during the current LHC shutdown.

Intensity Frontier experiments have relatively modest computing needs, at least in comparison to those of the Energy Frontier experiments.  Any single such experiment is expected to produce ``only'' a petabyte of data over its entire lifetime, compared to CMS or ATLAS which will produce several petabytes per year.  Thus it should not be difficult to provide the needed scale of computing resources for these experiments as long as sufficient funds are available.  These experiments too will be able to help themselves by actively pursuing opportunistic resources and operational efficiencies as the Energy Frontier experiments are.

Cosmic Frontier experiments have well-defined storage needs, and these become competitive with those of CMS and ATLAS in future years.  The Dark Energy Survey should produce ``only'' a petabyte of data by 2016 (well within current capabilities), but LSST could produce 100 PB by 2030.  The Square Kilometer Array could produce as much as 1500 PB/year when it is operational in the 2020's.  In addition, these experiments could have very different access and processing patterns than those of the accelerator-based experiments.  

The large increase in survey data means that statistical noise will no longer determine the accuracy to which cosmological parameters are 
measured. The control and correction of systematic uncertainties will determine the scientific impact of any cosmological survey. 
Achieving the goals of current and planned experiments will, therefore, require the processing and analysis of experimental data streams, 
the development of techniques and algorithms for identifying cosmological signatures and for mitigating systematic uncertainties, 
 and detailed cosmological simulations for use in interpreting systematic uncertainties within these data. 

There are three primary computational tasks associated with sky surveys: image generation, image processing, and cosmological simulation. The first 
is primarily an HTC task, the second is alredy running in HPC mode using up to thousands of processors, and the third uses cutting-edge HPC applications.  The computing requirements for image simulation and image processing will increase greatly over the next five years and, while substantial
on the order of 100 million hour, are expected to be accomodated at DOE and NSF centers. The HPC hours needed for cosmological simulations are extreme, however, reaching 10s of billions of hours by 2017~\cite{bib:NERSCreport}. The current outlook
makes it unlikely that HPC centers have adequate capacity to meeting these needs for cosmological
simulations on this time scale. 
 
Likewise, researchers performing Lattice QCD theory calculations 
-- which are essential for interpretation of many experiments done in high-energy and nuclear physics -- 
face an expected deficit in computing cycles. In in a recent report~\cite{bib:NERSCreport} LQCD researchers estimate needing 10s of billions of  hours in 2017, much more than will be available 
for LQCD if historical trends hold.

HPC needs for accelerator research and design are growing, too, but are expected to be accommodated by planned HPC center capacity and capability increases~\cite{bib:NERSCreport}.

\section{Will Distributed Computing Models Be Adequate?}
\label{sec:grid}
Because of their unique scales, particle physics experiments require unique computing solutions.  A modern experimental collaboration such as the ATLAS or CMS experiments includes thousands of scientists who are truly distributed over the whole world, spanning all the easily habitable continents.  They all need access to computing resources to perform their work and make scientific discoveries.  Meanwhile, the LHC experiments produce petabyte-scale datasets each year, and millions of CPU hours are needed for the processing and analysis of the data.  Historically, all of the computing resources for an experiment were hosted by the laboratory that operated the experiment, with some smaller installations at collaborating institutes.  However, this model has become less feasible as both experimental collaborations and recorded datasets have grown.  The necessary resources have large aggregate power and cooling demands, making it difficult to operate them all at one site.  While communities of researchers are generally willing to invest in the needed computing resources, they are reluctant to place them at a remote site.  Thus, a more distributed solution is necessary.

The current solution for distributed computing is grid computing.  A simple definition of a computing grid is an infrastructure that allows different administrative domains to share access to services, processors and storage with a select set of users and groups of users.  It is assumed that each organization participating in the grid provides its own computing resources.  The processors and storage can be distributed over a very wide geographic area, and are typically organized as clusters of computers that accept processing jobs through a batch system.  The computing clusters that are members of a grid provide a uniform environment for user batch jobs, even if each cluster has some unique local configuration.  As a result, any user job can, in principle, be executed at any site on the grid with little user customization required.

This fits the paradigm of experimental particle physics especially well.  The paradigm is that of high-throughput computing (as opposed to high-performance computing); most of the computational problems are embarrassingly parallel, as each data event recorded is statistically independent of the others.  Any data-processing task can thus be distributed across many batch jobs and many grid-computing sites, with the results being straightforwardly combined when all of the processing is done.

Distributing the computing resources across many locations has a number of advantages over consolidating them at one site.  It is possible to leverage local infrastructure and local expertise at each site, leading to greater engagement in the projects by individual institutions.  Computing clusters at sites such as universities are more likely to be multi-purpose, with many different researchers and scientific domains participating.  This allows greater resource sharing -- the peak processing time for one scientific domain may be different for that of another, and thus an active project can make opportunistic use of the resources provided by a temporarily less active project.  While each community does bring its own resources to the grid, it has access to the resources of others.  The ultimate result is a larger amount of computing available for any scientist's peak needs.  The tradeoff for this increased computing power is the challenge of managing a computing facility that is distributed over many different organizations, cultures and time zones, and of maintaining good throughput for users over such a system.

A computing grid has several ingredients.  There must be a set of individual computing facilities, each of which maintains some sort of batch-job submission system for scheduling access to particular resources in the cluster.  The resources in question are typically CPU's, but in the future one could imagine the scheduling of memory or network bandwidth.  The specific scheduler used at a given facility is irrelevant.  The facility is made available on the grid through a ``computing element'' (CE) that serves as a gateway for remote job submission.  It provides a uniform interface to the heterogeneous systems at each site, so that users wishing to use the site don't need to know the underlying details for job submission.

For cybersecurity purposes, this gateway needs to verify that a given user is allowed to use the local resources.  Thus, an elaborate system of user credentialing is needed.  Each user is a member of a ``virtual organization'' (VO) that validate user identities and issue credentials known as ``grid certificates'' that can be attached to a batch job when it is submitted to a grid resource.  These credentials allow for the tracking of batch jobs back to individual users.  In an HEP context, a VO is typically comprised of the participants of a single experiment.  Each computing site defines which VOs it is willing to accept jobs from, and often the site will need to implement a particular environment that the VO needs for its jobs to execute correctly, such as the software base of a given experiment.  When a user job is presented to a CE, the CE examines the grid certificate, decides whether the job will be allowed to run, and assigns it to the appropriate batch queue on the basis of the user identity.

Users need tools to submit jobs to grid sites.  These tools are known as middleware and are developed and maintained by the grid operators.  Users can use tools such as HTCondor and Bosco to submit jobs from their desktop computers.  HEP experiments usually provide some sort of wrapper around the middleware that serves as an interface with other services, such as data catalogues.  One growing means of job submission is through pilot systems.  In such a system, several computers serve as a factory for pilot jobs that are submitted to many grid sites.  The initial task of a pilot job is simply to hold on to a batch slot at a site.  Each pilot job can last for a long time on the host machine.  After determining the available environment on the machine, the pilot job can request a user job from a central task queue.  This allows for good matching between the available environment and the available work, and also allows a VO to use the central task queue to set priorities across the entire grid (rather than only at individual sites).  Any pilot job can take on multiple user jobs during its lifetime.  This saves the trouble of each user job having to authenticate at a site; only the pilot job needs to be authenticated when it begins.

Data storage and management is a challenge for grid systems.  While batch jobs can come and go and easily return resources to the grid for use by others, data tends to be more static.  What grid sites a given job can run on is often limited by the need to have a particular set of data files at the site.  This situation is evolving through the development of federated storage systems, in which files at one site can be made available for reading at another site straightforwardly.  For placing data at a site, there are tools such as SRM that provide a generic interface to the heterogeneous storage systems available at sites.

The success of grid systems depends on robust wide-area networking for both the submission of the job (which typically requires a small data transfer) and the retrieval of job results (which sometimes requires a much larger data transfer, depending on the nature of the output).  In addition, any infrastructure of such a scale needs good monitoring and accounting systems, to make sure that the grid is continually functioning and to understand usage patterns.

Different grids are operated by different national or regional entities.  These grids have grown organically, each with their own middleware that serves as the interface to the computing resources.  In the United States, the grid infrastructure used for particle physics is operated by the Open Science Grid (OSG).  The OSG provides much of the infrastructure needed for particle-physics experiments to run their workflows, such as the grid middleware, and supplies operational support through its Grid Operations Center.  Because of the need for a US-based grid infrastructure, the US HEP community helped develop and launch the OSG and continues to derive great benefit from it.  The OSG currently has about 120 affiliated sites and supports about 30 VOs.  

\begin{figure}[h]
\includegraphics[width=\textwidth]{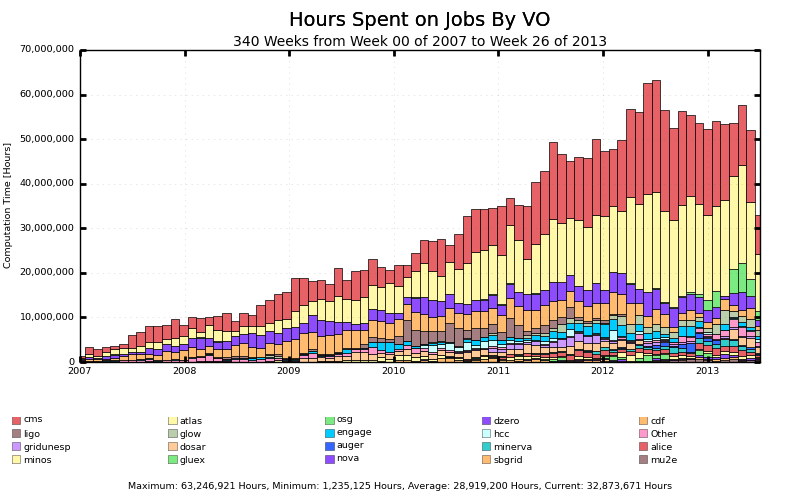}
\caption{Growth in OSG usage over the past six and half years.}
\label{fig:OSG-growth}
\end{figure}

Figure \ref{fig:OSG-growth} shows the steady growth in OSG usage over the past six and a half years.  While most of the activity and most of the growth has come from the LHC experiments (as discussed below), a large number of VOs have made substantial use of the grid, and the grid computing community continues to diversify.  Science areas such as protein structure studies, anti-earthquake engineering, and brain research have made excellent use of OSG resources.  This makes the OSG an excellent example of a project started by particle physicists that has had broader impact on the general scientific community. 

For Energy Frontier experiments such as CMS and ATLAS, with aggregate datasets in the tens of petabytes, the grid is the default location for most of the computing activity.  The LHC experiments are unusual in the history of particle physics in that it was planned from the very beginning that most of the computing activity would take place away from the host laboratory.  About 5/6 of the total CPU and disk available to the experiments is not at CERN but at distributed sites.  The CERN facility is only responsible for a first-pass reconstruction of the data shortly after they are recorded, plus some calibration and alignment activities.  Everything else required to perform a physics measurement -- re-processing with improved calibration and alignment, skimming to produce datasets more concentrated in events of interest, Monte Carlo simulation production and archiving, and final-stage user data analysis -- is conducted over the grid.  Most of the workflows described above are centrally managed by operations teams who decide which grid resources are used for which workflows.  The exception to this is user analysis, which is in the hands of individual physicists.  Historically, users have been constrained to submit their jobs to grid sites have a copy of the data sample that they wish to analyze.  However, this constraint has recently been relaxed through the deployment of Xrootd data access over the wide-area network, as discussed below.

Figure \ref{fig:OSG-usage} below shows OSG usage by virtual organization over the past year.  We see that ATLAS and CMS dominate the OSG, using about six million CPU hours per week.  Contrariwise, it should be noted that US-based grid computing resources play an outsized role in LHC computing, relative to the US-based fraction of the experimental collaborations.  For instance, on CMS the US grid sites provided about 40\% of the CPU time over the past year, while only about 1/3 of CMS membership is from the US.  The equivalent numbers for ATLAS are about 30\% of the CPU time and about 20\% of the collaboration. Besides being the biggest customer of CPU time, the LHC experiments are also the largest provider of processing resources, which can be opportunistically available to other VOs.  Thus we can see that the LHC experiments, the Open Science Grid and the US LHC users are all interdependent on each other for their success.

\begin{figure}[h]
\includegraphics[width=\textwidth]{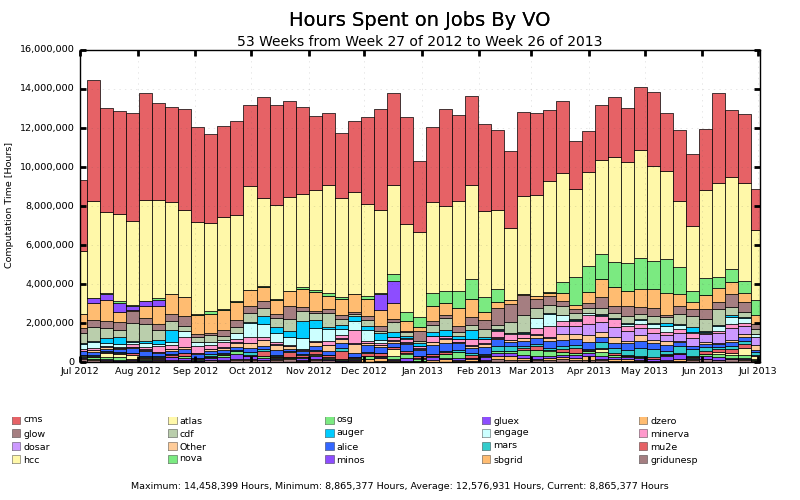}
\caption{OSG usage by vitual organization.}
\label{fig:OSG-usage}
\end{figure}

Since the data processing and analysis of the LHC experiments is dependent on grid computing, it is fair to say that grid computing is a key technology for discovery in particle physics.  Looking back to the Higgs boson observation of July 2012, we can see that ATLAS and CMS were able to turn around their data analyses very quickly.  Events recorded just a few weeks beforehand were included in the analysis.  The significant computing resources available through the WLCG, and their efficient and straightforward usage, made it possible to analyze this data quickly, to compute the significance of the observations and the strength of limits, and (perhaps most importantly) to prepare the necessary simulation samples that were needed to understand the data.  In his remarks at the end of the CERN seminar announcing the Higgs observation, CERN Director General Rolf Heuer explicitly recognized the role of the WLCG in the work.

At the very least, since the LHC experiments plan on relying on grid computing resources for the foreseeable future, we can anticipate that any future LHC discoveries will depend on the success of distributed computing.  While exactly how distributed computing is used may change in the future (as discussed by the Energy Frontier group), and while grid computing may not always be its implementation, we expect that distributed computing is here to stay, as the model has proven to be successful and it will be a great challenge to concentrate all computing resources for a single experiment in one location in the future.  It will be important to sustain and further develop our distributed-computing infrastructures.

\subsection{Research and development in grid infrastructures}

While grid computing has been a very successful technology for modern particle-physics experiments, especially those at the LHC, our operational experience has exposed a number of issues that should be addressed in the future if grids are to continue to serve the needs of the field.  Many of these can be addressed with either small-scale R\&D efforts, or through the promotion of best practices that are already known but not widely adopted.  In the end, none of these issues should be seen as show-stoppers for the long-term growth of distributed computing, but these developments should be pursued to improve efficiency and ease of use.

\subsubsection{Identity management}

The need for grid certificates has been the bete noir for many a particle physicist.  These certificates are used for identity management, and the process of obtaining a certificate can be quite convoluted, possibly requiring the use of multiple Web browsers and a number of utility programs to put the issued credentials in a form that can be used when a grid job is submitted.  Indeed, the Snowmass physics studies that were done on the grid brought many new users (mostly theorists) into the grid system, and they resisted the process of obtaining certificates.

However, just about every university campus already has some form of single sign-on that can be used to validate a user's identity for campus computing resources; campuses could then vouch for the identity of their users to the outside world.  The CILogon service (http://cilogon.org) uses this campus-level vetting to generate grid certificates through a simple Web-page username and password login.  A number of grid computing sites will accept these certificates, but their usage needs to become more widespread.

Of course a better solution still would be to eliminate the use of certificates altogether, or to at least make them invisible to ordinary users.  This can be achieved through pilot-based submission systems, where it is the pilot jobs that are authenticated rather than the user jobs that the pilots subsequently execute.  But the pilot jobs must still be able to provide traceability back to the user whose job is being run, so that sites can identify users if the need arises.  The OSG is currently studying the matter of traceability.  The use of pilot jobs is discussed further below.

\subsubsection{Streamlining operations}

As it currently stands, the operation of the computing grid for particle-physics experiments is very labor intensive; it is not very different from the operation of a complete multi-purpose particle detector.  Human effort is required nearly 24/7, with multiple people monitoring multiple layers of the system.  Many of the people who do this work are non-expert physicists in the collaboration who serve as computing shifters.  They observe symptoms of a problem at a particular computing site, but the root cause may not be a site problem; experts must intervene to untangle what is going on.  A large contingent of full-time experts are regularly busy solving infrastructure problems and optimizing resource usage.  CMS has 30 FTE (from 60 people) working on computing operations, while ATLAS operates with a somewhat smaller team. This is in addition to the teams of people who are actually operating the grid infrastructures themselves, which is about 10 FTE for the OSG.  As in so many cases, staffing costs are a significant element of the computing budgets for the experiments and the grids.

How can all of these operational tasks be performed with fewer people?  We should look to develop self-learning systems that can monitor the infrastructure and the computational work to be done, figure out the resource requirements for the work, and optimize the scheduling of the work.  These systems could identify problems on the infrastructure and make sure that work isn't scheduled into those problems.  Related to this is a need for self-healing systems that can solve the problems themselves, or at least flag them for intervention by a human.  Research in this sort of machine learning might be funded through DOE ASCR or NSF OCI programs, and we should not hesitate to look towards industry for solutions, as Internet companies are also running very large distributed computing infrastructures.

\subsubsection{Storage management}

Batch jobs are naturally suited for the grid -- each job is short-lived and when it is finished, the processing resource is freed up and can be used by a different job.  At the moment, the grid is not as well suited for data storage.  Stored data does not clean itself up when it has outlived its usefulness, and indeed users tend to be paranoid about holding on to obsolete data.  How to store and access data in the grid environment is perhaps the largest unsolved problem of grid computing.

Storage management has two main forms: the management of large-scale datasets that are of use by large groups within an experimental collaboration, and the management of small-scale datasets owned by particular users.  The former issue is discussed by the Storage and Data Management group; we only note here that there is some debate within the community whether there will be a scaling problem or not as LHC datasets grow.  In any case, it is expected that experiments will continue to take responsibility for developing management systems for their own data.

Small-scale user storage is a more general issue for grid infrastructures, as many communities beyond HEP want to manage small (compared to HEP) datasets.  Over the past few years, we have seen the emergence of the ``dropbox'' model, in which a particular set of files can be placed ``in the cloud'' and then made accessible on any computer.  The file set can be managed by each user in the way he or she desires, with a set of straightforward API's.  It will be useful for distributed-computing infrastructures to develop their own dropboxes, so that users can have access to their data anytime and anywhere in the system.  The first forms of such global data access have now emerged in the form of CMS's ``Any Data, Anytime, Anywhere'' (AAA) project and ATLAS's ``Federating ATLAS storage systems using Xrootd'' (FAX) project, both of which rely on Xrootd as their underlying technology. [add references] These systems are already giving experiments and individual users greater flexibility in how and where they run their workflows by making data more globally available for access.  Potential issues with bandwidth can be solved through optimization and prioritization.  The further development of such systems should be encouraged.

\subsubsection{Simplification and scaling of job submission}

As the grid currently works, each new job that starts in a batch slot must be authenticated by a site's grid gatekeeper by checking the user's credentials.  Given that a user may submit hundreds or even thousands of jobs at time, the same credentials are checked over and over again.  This is obviously an inefficient process.  One solution that is already being deployed is the use of pilot jobs.  A pilot job can be authenticated once at a site and then pull in and execute multiple individual user jobs in sequence from an external task queue.  In this case, the validation of the credentials of the individual jobs must fall to the pilot system.  A drawback to this scheme that needs further exploration is that it can be difficult to trace the origins of a given job submitted through a pilot system back to the user who submitted it.  Solutions such as glexec exist, but these need more widespread deployment and testing.

\subsubsection{Dynamic scheduling}

Historically, the only resource that a batch job needed to have scheduled for it was a CPU.  Other resources were more static in nature and could be used (or not used) at any time without concern.  However, batch jobs are now requiring a more diverse set of resources.  Some jobs may have special memory requirements or wish to use the wide-area network in a real-time fashion.  As a result, job scheduling mechanisms will have to be able to schedule a more heterogeneous set of resources.  HTCondor is already starting to move in this direction; these efforts should continue to be supported.

\subsubsection{Clouds vs. grids}

The cloud computing paradigm presents users with a virtual machine that requires extensive configuration by the user.  In contrast, grid computing sites usually have the appropriate environment already set up.  The largest cloud-computing providers are commercial entities, such Amazon, which charge for the use of CPU time, storage, and inbound and outbound network bandwidth.  To date, there have been several small experiments in using cloud facilities for HEP applications.  CMS has configured machines from the Amazon cloud to appear as an extension of an existing Tier-2 cluster, running the jobs on the cloud but reading data from dedicated CMS sites with the Xrootd protocol.  ATLAS has performed a number of large workflows with both Amazon and Google clouds, some of which was paid for with a grant from Google.  These tests used from hundreds to thousands of CPU’s and ran from one to six months in duration.

While these tests were quite successful, we do not believe that commercial clouds are currently a viable option for large-scale particle-physics computing.  This is largely due to costs; it is currently cheaper to operate a Tier-2 center at a university rather than to buy equivalent cloud resources.  Costs are a particular concern with respect to data, as the cost of hosting data in the Amazon cloud is \$0.10/GB/month.  Data can be read in over the WAN at no charge, but transferring data out to a user’s home computer can cost as much as \$0.12/GB.   The output files must be staged out over the network, which also has a cost.  In addition, the computing sites operated for HEP often have specially-tuned infrastructures that support the particular needs of these computational tasks and that are not readily available for generic cloud resources.

That being said, the costs of commercial clouds could well fall over time, and we expect that more academic computing centers will be setting up cloud interfaces to their clusters.  Given how rapidly clouds have emerged in commercial markets, the development of opportunistic clouds could come quickly.  Thus we should still develop the capabilities to fully use computing clouds, should conditions become more economically favorable in the future.

\subsection{Conclusions on grid infrastructures}

The fundamental paradigm of using grid computing for the embarrassingly-parallel problem of HEP event data processing seems to be sound.  The distributed computing approach should be able to scale up appropriately, especially with the developments that have been envisioned and in some cases are currently underway.  Many of the issues are about how to make the grid easier to use, both for users and for operators, more than about any fundamental issues of scale.

There has been a major push to involve a wide range of sciences in grid computing, with some notable successes in non-HEP science domains.  But HEP has always been, and still is, by far the largest consumer of grid resources, as seen in Figure~\ref{fig:OSG-usage}.  If we believe that distributed computing will remain a cornerstone of computing for particle physics, we must be prepared to take the lead on investing in developing this technology; we cannot assume that anyone else will step up and do it for us.

\section{What role will national computing centers play in computations for HEP?}

The U.S. national HPC centers run by the Department of Energy Office of Science and the National Science Foundation have traditionally served the particle physics and high energy physics (HEP) communities by enabling and supporting large simulations in Lattice Quantum Chromodynamics (LQCD), accelerator research and design, cosmology, and supernova physics. The astrophysics community has also used HPC centers to produce mock catalogs in support of satellite and ground-based sky surveys.
 
Over the next fine to ten years HPC centers are expected to continue their roles with these traditional communities. However, the computational and storage resources needed to support upcoming initiatives in part sponsored by the Department of Energy Office of Science -- including the LHC upgrade and  the Large Synoptic Sky Survey -- will stress the ability of HPC centers, even if they are able to continue the traditional HPC growth rate of 1.8-2.0 X per year.

\begin{figure}[h]
\includegraphics[width=\textwidth]{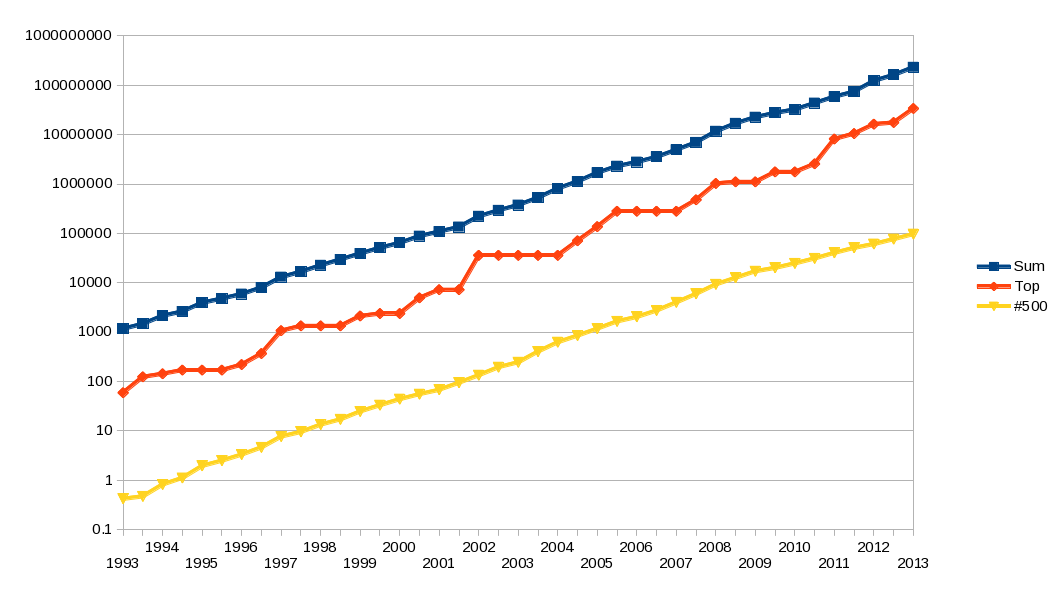}
\caption{Performance of the Top 500 supercomputers as compiled by Top500.org.  Computational power doubles about every 14 months. An exascale system in the U.S. may be achieved sometime around 2020-2023, depending on funding and the rate of technology advances needed to control power consumed by such a system.}
\label{fig:Top-500}
\end{figure}

The National Energy Research Scientific Computing Center (NERSC)-- the production HPC center for the U.S. Department of Energy's Office of Science -- has conducted two reviews with the HEP community to determine their computing and storage needs. The first review, in 2009, targeted needs for 2014, while the second, in 2012, targeted 2017.  The findings reveal a shortage of anticipated resources needed to support DOE HEP mission needs. 
The shortage is largely driven by the LQCD and cosmological simulation communities.

In addition to the resources needed by the traditional HEP HPC community, funding agencies 
-- in particular the DOE HEP office -- are becoming increasingly cognizant of the computing 
needs of all sections of the HEP research community. In part this is due to the awareness 
by the traditional HTC communities that they too find HPC center resources valuable. 
Some of these groups -- most notably HEP theorists carrying out Monte Carlo simulations that 
are used by experimentalists -- are beginning to use HPC centers(1) \textcolor{red}{Needs a reference?}. 
For example, the DOE Offices of HEP and Advanced Scientific Computing (ASCR) have started a joint partnership to transform the GEANT-4 code into an efficient parallel code that can take advantage of multicore and HPC platforms. In parallel, the Energy Frontier experimental community and major Cosmic Frontier experiments are preparing to use HPC facilities.
 
The upcoming LHC run at a center-of-mass energy near 14 TeV will strain the computing capacity of the ATLAS and CMS experiments, as described earlier. The higher energies, luminosities and trigger rates expected in the LHC high energy run starting in 2015 require substantial increases in computing capacity to handle them, and given the end of Moore's Law scaling it will be difficult for computing resources to keep pace with demand by simply maintaining a flat computing budget.  While HEP experiment data rates have traditionally been limited by trigger bandwidth and the throughput of DAQ systems, they are now limited by the available computing capacity (both for processing and storage), and scientific results are in some case limited by the rate at which simulated samples can be created.

An attractive solution to the latter problem of simulation is to port elements of ATLAS and CMS software (Monte Carlo event generators such as Alpgen and simulation of events using the Geant4 toolkit appears to be the best starting points) to run on HPC resources that are at existing DOE resources, opening up the availability of supercomputers for what have traditionally been HTC applications.  ALTAS is already developing a front-end cluster to allow it to interact with these computers via the Open Science Grid.
 
Using supercomputers for HEP experiments such as ATLAS and CMS has these benefits:
\begin{enumerate} 
\item{By increasing the resources for simulation, the discovery reach of the experiment is substantially increased.}
\item{Because leadership computers have very large capacity and is a shared resource across DOE, the experiments will be better able accommodate large fluctuations in resource needs typical of HEP experiments.}
\item{The availability of supercomputing resources to HEP physicists will speed up the development of software that takes advantage of new computation architectures in a way that is broadly applicable to the new generation of commodity hardware.}
\end{enumerate}
 
HPC centers have a number of additional attractive features.

\begin{itemize} 
\item{They can play a facilitating role in helping particle physicists make the transition to new computing architectures by providing access to prototype architectures, HPC consulting, and training classes.}
\item{The facilities are run by computing experts, with 24x7 monitoring and response to problems}
\item{Centralized problem tracking and solutions}
\item{Consulting and account support}
\item{Centralized repositories of software and data}
\item{A traditional sustained growth rate of resources (~2X/year computing, 1.7X/year storage)}
\item{State of the art networking (e.g., NERSC has advanced, secure, and open network access; all access to NERSC now goes through a state-of-the-art 100 G network that includes advanced security monitoring and completely open and transparent access for science teams. ).}
\item{State of the art hardware and software technologies}
\item{Access to the large storage, high I/O capabilities, and networking are available.  Access to computational resources ``close'' to the data to be analyzed reduces the need move large data sets many times over the network.}
\end{itemize} 

From the standpoint of the HPC centers, there are challenges to accommodating an expanded workload. The addition of traditional HTC workflows, and those currently using the WLCG in particular, onto 
national HPC centers will require integrating them into job scheduling systems already in place. 
A workload containing many serial compute jobs, which characterize most of WLCG computing, 
is not a good fit with large parallel computational systems with expensive node interconnects. 
Identity management and data ownership and management pose other obstacles. Finally, providing 
computing, storage, and services to a new community requires an additional resources from 
funding agencies either through the normal resource allocations processes or through outside 
funding from the community to augment the HPC centers' base programs.
 
Meeting these challenges is not completely new to HPC centers. For exanmple,
NERSC has a long history of supporting ``non-traditional HPC'' HEP research.  
NERSC hosts the PDSF cluster (which has its origins in the Superconducting Supercollider), 
which is the Tier 1 site for the Daya Bay neutrino experiment and 
a Tier 3 site for ATLAS. PDSF is also currently used by ALICE, STAR, and CUORE. 
 NERSC also hosts the Lattice Connection and Deep Sky web data-driven science gateways for 
LQCD field configurations  and supernova discovery. 
NERSC's HPC systems were used to calculate the more than 100,000 simulated realizations of the 
early universe needed to process results from the Planck CMB satellite mission.

\subsection{HPC Center Growth Rates and Plans}
 
As shown in Figure \ref{fig:NERSC-Computational-Hours} above the projected need for HEP production computing within DOE will outpace even the traditional Moore's Law increase of approximately 2X per year.  The red line in the plot connects the stated requirements with 2012 usage and shows that just the traditional HPC HEP computing community has needs that far exceed the traditional trend lines. 
 
We are at a critical time for HPC computing. To continue the historic, Moore's Law-like, growth in HPC computing power, the HPC community must adapt to build (with HPC vendors) and use new energy-efficient technologies. The path forward will require use of low-power, simple fundamental processing units. The current leading architectural candidates are systems that incorporate NVIDIA GPUs (Graphical Processing Units), the Intel Phi ``many-core'' processor, processors based on AMD's APU technology, and units integrating ARM-based technology.

This pressing need for resources and innovation comes at a time of uncertain budgetary forecasts. The DOE exascale initiative targets enabling an exascale compute system by 2022-2023 -- with the actual timeframe dependent on funding and technology advances -- through the Fast Forward (processor design) and Design Forward (system design) partnerships with HPC vendors. The Exascale Initiative includes partnerships with government, the computer industry, DOE laboratories, and academia.  Goals include
\begin{itemize}
\item{1,000 times more performance than a Petaflop system}
\item{Capability to execute 1 billion degrees of concurrency}
\item{20 MW power consumption or less}
\item{200 cabinets or less (space constraint)}
\item{Development and execution time productivity improvements}
\end{itemize}
 
[Other plans? EXCEDE / NSF .]

While the past focus on HPC centers has largely been on massive parallel computations, the need for, and ability to facilitate handling of, accompanying data has quietly grown over the years. 
All the DOE Office of Science HPC centers have drafted plans to accomodate data-driven science and 
there is a proposal for a Virtual Data Center with DOE to address data issues. 

[Other plans? NSF?]

\section{What coordination will be required across facilities and research teams, and are new models of computing required for it?}

The scale of computing facilities and resources has greatly increased over the past few decades of HEP research, as the size of experimental datasets and the sophistication of theoretical calculations and simulations has grown.  It has reached the point where large organizations are now needed to interact with the computing providers in order to get the most use out of them.  A fair amount of expertise is required to design, optimize and then operate the workflows on the most modern infrastructures.

Larger experiments, like those at the LHC, do have the resources to make full use of infrastructures such as that of grid computing.  But many smaller projects simply lack the available effort to do so.  This seems to be especially true for the projects in the Intensity Frontier area, which typically have many fewer physicists working on them compared to CMS or ATLAS.  This is a clear case where additional coordination between the research teams could be beneficial.  Fermilab, as the host laboratory for many of these experiments, is already trying to play a coordinating role, at least at the level of providing uniform event-processing frameworks across many experiments.  This model should be extended even further, to help the Intensity Frontier projects have a uniform approach to computing that can take advantage of the efficiencies of scale that come from working together.  We suggest that an Intensity Frontier Computing Consortium could be formed to advocate for the computing needs of the Intensity Frontier experiments.  Given that the funding for research at the Intensity Frontier is coordinated through one program at DOE HEP, it might be possible to find the funds to help support the computing efforts of multiple experiments there.

As the quest for computing resources takes researchers away from their current ``natural'' computing homes of distributed HTC on the grid and HPC at specialized shared facilities, it will become more important for users to be able to move between the different computing paradigms easily.  The ability to make seamless transitions will rely on work that is both technical, such as creating uniform systems of identity management and job submission, and political, such as coordinating between entities such as the OSG, XSEDE, and national centers such as NERSC.  It would be useful for the leaders of those organizations to discuss how they could start to build a computing infrastructure that can truly serve all the members of the HEP community.

Engineering and technology advancements will enable energy-efficient, performant processors and systems, but the 
programming paradigms employed by the application developer will likely become more complex, for both HPC and HTC 
applications.   
Code development on these new architectures will pose a serious challenge. 
The programming strategy, in practice since the late 1990s, of optimizing code to minimize CPU cycles and 
inter-node communication must be replaced with a paradigm that stresses minimizing data movement at all levels 
(from on-chip cache, to on-node shared memory, to storage media) and maximizing SIMD (vector) 
processing operating on contiguous data, and use of light-weight threads of execution. 
Most HPC software may have to be modified significantly or rewritten to execute acceptably well 
(or at all in some cases) on these architectures. Programming models for their architectures are 
nascent and/or proprietary to a particular vendor's architecture. Some science communities have 
embraced one technology or another, while others are waiting to see which programming model 
and/or language ``wins out.'' In either case, a significant investment in code development will be required.
 
This issue is relevant to those applications that currently use the WLCG. 
Although traditional multi-core processors and systems will continue to exist for some time, 
growth in computing power will come from ``massively multi-core'' processors. If the HEP community does not adapt codes 
to run in parallel (using fine-grained parallelism ({\it e.g.} GPUs) and possibly 
coarse-grained parallelism ({\it e.g.} MPI)) it will be precluding itself from using a vast source 
of computational power in a time of greater need.
 
\section{Summary}
\label{sec:comp-summary}
Powerful distributed computing and robust facility infrastructures are essential to the continued progress of particle physics across the Energy, Intensity, and Cosmic Frontiers.  Those approaches, and the theoretical work needed to support them, require a combination of both HTC and HPC systems.  The dominant consumers of HTC are the LHC experiments, who have been well served by it and should be in the future also.  Most Intensity Frontier experiments can be supported by HTC also.  HPC is needed for applications such as lattice QCD, accelerator design and R\&D, data analysis and synthetic maps, N-body and hydro-cosmology simulations, supernova modeling, and more recently perturbative QCD.  HPC tasks have historically been carried out at national centers that have focused primarily on HPC facilities, but these centers have begun to embrace the problems that are addressed by HTC and are interested in attracting scientists who want to work at these centers.

Energy Frontier experiments face a growth in data that will make it a challenge to supply the needed computing resources.  Doing so is possible as long as certain conditions are in place and specific steps are taken to keep up with evolving technologies.  It requires near-constant funding of the WLCG, greater efficiencies in resource usage, and the evolution of software to take advantage of multicore processor architectures.  These experiments should also pursue and take advantage of opportunistic resources, be they in commercial clouds (which are not currently viable as purchased resources), universities, DOE centers or elsewhere.  The experiments would also benefit from further engagement with national HPC centers, which have resources available to HEP experiments and can support efforts to make use of HPC systems that have not traditionally been used for HTC applications for applications such as detector simulations.

Intensity Frontier experiments have smaller computing needs in comparison, and there is nothing technically that prevents them from being met.  Such experiments should be aware of the existence of resources available to them through the OSG or at national computing centers, and they could benefit from a coordinated effort amongst them to gain access to resources and share software and training.

Cosmic Frontier experiments and the simulations needed to interpret them are among the drivers of a need for growth in HPC resources in the coming years, along with lattice QCD and accelerator design.  Indeed, demand for access to HPC across HEP is expected to exceed the expected availability.  
Such computations are critically needed to interpret results from a number of important experiments and realize scientific returns from national investments in those experiments.  The NERSC report on HEP computing needs indicates a shortage of HPC resources needed for HEP by a factor of four by 2017.  While funding and technology development needed to sustain traditional HPC growth rates are uncertain, they must be maintained to support HEP science.

Distributed computing infrastructures, which have been critical to the success of the LHC experiments, should continue to be able to serve these and other applications even as they grow in scale.  There are no show-stoppers seen in increasing scale, but various developments should be purused to improve efficiency and ease of use.  Keeping sufficient staff support at a reasonable cost is a continuing concern; finding operational efficiencies could help address this.  Given that HEP is the largest user of distributed scientific computing, currently in the form of grid computing, members of the field must continue to take a leadership role in its development.

National centers play an important role in some aspects of computing, and HEP might be able to take advantage of an expanded role.  It is already used in many of the applications listed above.  While there are not enough resources dedicated to HEP available at the centers to rival those of the WLCG, experiments should explore the use of the centers as part of their efforts to diversify their computing architectures.  These centers do have access to large, state-of-the-art resources, operational support and expertise in many areas of computing.

We expect that distributed computing and facility infrastructures will continue to play a vital role in enabling discovery science.

%% file: authorlist.tex


\begin{center}
{\large Conveners: Kenneth Bloom$^1$,  Richard Gerber$^2$}\\
\bigskip
$^1${\it Department of Physics and Astronomy, University of Nebraska-Lincoln}\\
$^2${\it National Energy Research Scientific Computing Center (NERSC), Lawrence Berkeley National Laboratory}\\
\end{center}
